\documentclass[structabstract]{aa}
\usepackage{graphicx}
\usepackage{txfonts}
\usepackage{upgreek}
\usepackage{tipa}
\usepackage{verbatim}
\usepackage{natbib}
\usepackage{color}
\begin{document}

\newcommand{\hoA}{o-H$_2^{18}$O 1$_{10}$-1$_{01}$}
\newcommand{\hoB}{o-H$_2^{17}$O 1$_{10}$-1$_{01}$}
\newcommand{\hoC}{o-H$_2$O 1$_{10}$-1$_{01}$}
\newcommand{\hoD}{p-H$_2$O 2$_{11}$-2$_{02}$}
\newcommand{\hoE}{p-H$_2$O 2$_{02}$-1$_{11}$}
\newcommand{\hoF}{p-H$_2^{18}$O 2$_{02}$-1$_{11}$}
\newcommand{\hoG}{o-H$_2^{18}$O 3$_{12}$-3$_{03}$}
\newcommand{\hoH}{o-H$_2$O 3$_{12}$-3$_{03}$}
\newcommand{\hoI}{p-H$_2^{18}$O 1$_{11}$-0$_{00}$}
\newcommand{\hoJ}{p-H$_2^{17}$O 1$_{11}$-0$_{00}$}
\newcommand{\hoK}{p-H$_2$O 1$_{11}$-0$_{00}$}
\newcommand{\hoL}{o-H$_2$O 2$_{21}$-2$_{12}$}
\newcommand{\hoM}{o-H$_2^{17}$O 2$_{12}$-1$_{01}$}
\newcommand{\hoN}{o-H$_2$O 2$_{12}$-1$_{01}$}
\newcommand{\hoO}{p-H$_2^{18}$O 3$_{13}$-3$_{20}$}
\newcommand{\kms}{km~s$^{-1}$}
\newcommand{\ms}{m~s$^{-1}$}
\newcommand{\cms}{cm~s$^{-1}$}
\newcommand{\halp}{H$\alpha$}
\newcommand{\msun}{$\rm M_{\odot}$}
\newcommand{\etal}{et~al.~}
\newcommand{\vsini}{$v~sin~i$}
\newcommand{\ctemp}{$^{\circ}$C}
\newcommand{\ktemp}{$^{\circ}$K}
\newcommand{\be}{\begin{equation}}
\newcommand{\ee}{\end{equation}}
\newcommand{\bd}{\begin{displaymath}}
\newcommand{\ed}{\end{displaymath}}
\newcommand{\bi}{\begin{itemize}}
\newcommand{\ei}{\end{itemize}}
\newcommand{\bfig}{\begin{figure}}
\newcommand{\efig}{\end{figure}}
\newcommand{\bc}{\begin{center}}
\newcommand{\ec}{\end{center}}
\newcommand{\hii}{{H\scriptsize{II}}}
\newcommand{\vlsr}{V$_{\mathrm{LSR}}$}
\newcommand{\vtur}{V$_{\textrm{\tiny{tur}}}$}
\newcommand{\vexp}{V$_{\textrm{\tiny{exp}}}$}
\newcommand{\vinfall}{V$_{\textrm{\tiny{infall}}}$}
\newcommand{\coa}{$^{12}\mathrm{CO}$}
\newcommand{\cob}{$^{13}\mathrm{CO}$}
\newcommand{\coc}{$\mathrm{C}^{18}\mathrm{O}$}
\newcommand{\lsun}{L$_{\odot}$~}
\newcommand{\lfir}{L$_{\textrm{\tiny{FIR}}}$}
\newcommand{\agua}{$X_{\textrm{\tiny{H$_2$O}}}$}
\newcommand{\ratioop}{o$/$p}
\newcommand{\ratiosept}{$X_{\textrm{\tiny{$^{18}$O$/$$^{17}$O}}}$}
\newcommand{\ratiohuit}{$X_{\textrm{\tiny{$^{16}$O$/$$^{18}$O}}}$}
\newcommand{\watersept}{H$_2^{17}$O}
\newcommand{\waterhuit}{H$_2^{18}$O}
\newcommand{\water}{H$_2^{16}$O}

\def\etal{et al.$\;$}


\def\kms{km\thinspace s$^{-1}$}
\def\Lsun{L$_\odot$}
\def\Msun{M$_\odot$}
\def\ms{m\thinspace s$^{-1}$}
\def\percc{cm$^{-3}$}

\title{The massive protostar W43-MM1 as seen by Herschel-HIFI water spectra: high turbulence and accretion luminosity\thanks{Herschel is an ESA space observatory with science instruments provided by European-led Principal Investigator consortia and with important participation from NASA.}}

\titlerunning{The massive protostar W43-MM1 as seen by Herschel-HIFI water spectra}

\subtitle{}

\author{F. Herpin,
	\inst{1,2}
	\and
	L. Chavarr\'{\i}a
	\inst{1,2}
	\and
	F. van der Tak
	\inst{3}
	\and
	F. Wyrowski
	\inst{4}
	\and
	E. F. van Dishoeck
	\inst{5,6}
	\and
	T. Jacq
	\inst{1,2}
	\and
	J. Braine
	\inst{1,2}
	\and
	A. Baudry
	\inst{1,2}
	\and
	S. Bontemps
	\inst{1,2}
	\and
	L. Kristensen
	\inst{5}
}

\institute{
Univ. Bordeaux, LAB, UMR 5804, F-33270, Floirac, France.
\and 
CNRS, LAB, UMR 5804, F-33270, Floirac, France, 
	\email{herpin$@$obs.u-bordeaux1.fr}
\and 
SRON Netherlands Institute for Space Research, PO Box 800, 9700AV, Groningen, The Netherlands
\and 
Max-Planck-Institut f\"ur Radioastronomie, Auf dem H\"ugel 69, 53121 Bonn, Germany
\and  
Leiden Observatory, Leiden University, PO Box 9513, 2300 RA, Leiden, The Netherlands
\and  
Max Planck Institut f\"ur Extraterrestrische Physik, Giessenbachstrasse 1, 85748 Garching, Germany
}

\date{accepted 2012 April 2 }
\abstract
   {}  
   {We present Herschel/HIFI observations of fourteen water lines in W43-MM1, a massive protostellar object in the luminous star cluster-forming region W43. We aim at placing this study in the more general context of high-mass star formation. The dynamics of such regions may either be the monolithic collapse of a turbulent core, or competitive accretion. Water turns out to be a particularly good tracer of the structure and kinematics of the inner regions, allowing an improved description of the physical structure of the massive protostar W43-MM1 and an estimation of the amount of water around it.}
   {We analyze the gas dynamics from the line profiles using Herschel-HIFI observations (WISH-KP) of fourteen far-IR water lines (\water, \watersept, H$_2^{18}$O), CS(11--10), and C$^{18}$O(9--8) lines, and using our modeling of the continuum spectral energy distribution. The spectral modeling tools allow us to estimate outflow, infall and turbulent velocities, and molecular abundances. We compare our results to previous studies of low-, intermediate-, and other high-mass objects. }
   {As for lower mass protostellar objects, the molecular line profiles are a mix of emission and absorption, and can be decomposed into 'medium'  (FWHM$\simeq$5-10 \kms), and 'broad' velocity components (FWHM$\simeq$20-35 \kms). The broad component is the outflow associated with protostars of all masses. Our modeling shows that the remainder of the water profiles can be well fitted by an infalling and passively heated envelope, with highly supersonic turbulence varying from 2.2 \kms~ in the inner region to 3.5 \kms~ in the outer envelope.  Also, W43-MM1 has a high accretion rate, between 4.0 $\times$ $10^{-4}$  and 4.0 $\times$ $10^{-2}$ \msun yr$^{-1}$, derived from the fast (0.4-2.9 \kms) infall observed.  We estimate a lower mass limit of gaseous water of 0.11 \msun\ and total water luminosity of 1.5 \lsun (in the 14 lines presented here). The central hot core is detected with a water abundance of 1.4 $\times10^{-4}$ while the water abundance for the outer envelope is 8 $\times10^{-8}$. The latter value is higher than in other sources, most likely related to the high turbulence and the micro-shocks created by its dissipation.}
   {Examining water lines of various energies, we find that the turbulent velocity increases with the distance to the center.  While not in clear disagreement with the competitive accretion scenario, this behavior is predicted by the turbulent core model. Moreover, the estimated accretion rate is high enough to overcome the expected radiation pressure.}
   

\maketitle

%

\section{Introduction}

After Hydrogen and Helium, Oxygen is the most abundant element in the Universe and water is one of the most abundant molecules. Due to the water in the Earth's atmosphere, space-based observations are necessary to study the vast majority of the water transitions.  Because of its importance for life and its sensitivity to dynamical, thermal, and chemical processes in the interstellar medium, the physics and chemistry of water is one of the main drivers of the Herschel Space Observatory mission \citep[hereafter Herschel,][]{Pilbratt2010} and particularly of the HIFI spectroscopy instrument \citep[][]{deGraauw2010}.  

Massive stars are rare but are the major contributors to the matter cycle in the Universe due to their short lifetimes and rapid ejection of enriched material. 
OB stars dominate the energy budget of star-forming galaxies and are visible at great distances.  
Their formation, however, is not well understood and the classical scheme for low-mass star formation cannot be applied as such to OB stars. Indeed, young OB stars and protostars strongly interact with the surrounding massive clouds and cores, leading to a complex and still not clearly defined evolutionary sequence. We generally identify in this sequence objects ranging from massive pre-stellar cores, high-mass protostellar objects (HMPOs), hot molecular cores (HMC), and finally to the more evolved UItra Compact HII regions stage where the central object begins to ionize the surrounding gas \citep[e.g.,][]{beuther2007b}. The classification adopted above is not unique but is consistent with the analysis of massive young stellar objects and HII regions in the Galaxy made by  \citet{mottram2011}. The most problematic issue in understanding the massive star formation process is to explain how to accumulate a large amount of mass infalling within a single entity despite radiation pressure. Models considering a protostar-disk system \citep[e.g.,][]{yorke2002, krumholz2005, banerjee2007} now quite successfully address how the accretion of matter overcomes radiative pressure. Actually, two main theoretical scenarios have been proposed to form high-mass stars, both requiring the presence of a disk and high accretion rates: (a) a turbulent core model with a monolithic collapse scenario \citep[][]{mckee2002, mckee2003}; (b) a highly dynamical competitive accretion model involving the formation of a cluster \citep[][]{bonnel2006}. In ionized HII regions, star formation could also be triggered by successive generations of stars \citep[e.g.,][]{deharveng2009}. One of the implications of the turbulent core model is that molecular line profiles should show supersonic turbulent widths while the competitive accretion produces cores which are subsonic, hence with rather small turbulence \citep[see][]{krumholz2009}.  
In the deeply embedded phase of star formation, it is only possible to trace the dynamics of gas through resolved emission-line profiles, such as obtained with HIFI.   

The Guaranteed-Time Key Program {\it Water In Star-forming regions with Herschel}  \citep[WISH,][]{vandishoeck2011} probes massive star formation through water observations using the HIFI and PACS \citep[][]{poglitsch2010} instruments. In particular, the dynamics of the central regions are characterized through the water lines and the amount of cooling they provide is measured.
In molecular clouds, water is mostly found as ice on dust grains, but at temperatures $T>$100 K the ice evaporates, increasing the gas-phase water abundance by several orders of magnitude \citep[][]{fraser2001, aikawa2008}. This suggests that water emission almost exclusively probes the warm inner regions of protostellar objects.  In order to collapse, the gas must be able, among several other conditions, to release enough thermal energy; a major WISH goal is to determine how much of the cooling of the warm region ($T>$100 K) is due to H$_2$O.  

The high velocity resolution of the HIFI instrument allows us to study the dynamics of the gas, detect outflows, and estimate infall and turbulent velocities present in protostellar envelopes.  Even though the water abundance in the envelope is small, low energy lines are highly absorbed by the envelope, producing a mixture of emission and absorption that requires the spectral resolution of HIFI in order to be studied. The high-energy lines are observed in emission and directly probe the warm regions since their lower energy levels are not excited in the envelope, and as such cannot absorb emission from the hot core. Thus, observations of a mixture of low- and high- lying lines of water and its isotopologues allow  a 'tomography' of the structure of protostellar envelopes. 


A first review of the WISH KP and results have been presented and compiled by \citet{vandishoeck2011} for a few objects only. We expect to confirm these results with the study of the whole sample. Interestingly, the water line profiles (transitions \hoK~ and \hoE) obtained in a low- \citep[NGC1333,][]{kristensen2010}, an intermediate- \citep[NGC7129,][]{johnstone2010}, and a high-mass \citep[W3IRS5,][]{chavarria2010} young stellar object (hereafter YSO) exhibit similar velocity components: a broad (FWHM $\sim$ 25 \kms) and a medium ($\sim$5-10 \kms) one. A narrower component ($<$5 \kms) is also observed, but not in the massive object, although it has been detected toward W3IRS5 by \citet{wampfler2011} in the OH emission. On the opposite, the water emissions differ in intensity, the massive objects being the strongest emitters. Another high-mass star forming region, DR21(Main) has been studied by \citet{vandertak2010} in the para ground-state water and $^{13}$CO lines. They derived a very low water abundance of a few $10^{-10}$ in the outer envelope while it is enhanced by nearly three orders of magnitude in the outflow (7 $\times 10^{-7}$). Water abundance estimates toward four other massive YSOs have been made by \citet{marseille2010} using the lowest two p-H$_2$O lines combined with the ground state p-\waterhuit~ line. Variations of the outer envelope water abundance are measured (from 5 $\times 10^{-10}$ to 4 $\times 10^{-8}$), but are not correlated with the luminosity or evolutionary stage. Finally, no infall has been firmly established so far for these objects.

In this paper we analyze the water observations towards the massive dense core W43-MM1 in the mini-starburst region W43, located near the end of the Galactic bar at a distance of 5.5 kpc  \citep[][]{motte2003, nguyen2011}. This region is one of the most interesting star formation regions in several aspects \citep[e.g.,][]{bally2010}. It has a star-forming rate that has increased by an order of magnitude compared to  $\sim 10^6$ years ago \citep[][]{nguyen2011}. 
The source is an IR-quiet dense core following the definition of \citet{motte2007} 
with bolometric luminosity 2.3 $\times 10^4$ \lsun and size of 0.25 pc. This massive dense core is expected to be fragmented on small scales and should be considered as a protocluster. W43-MM1 is the most massive dense core \citep[M$\sim$ 3600 \msun,][]{motte2003} of the four identified main millimeter fragments of which only MM1 is within the HIFI beam. The  mid-IR emission from the hot core is undetected because it is totally absorbed by the envelope, while the source is detected at 24 $\mu$m (peak flux= 0.3 Jy). A methanol and a water maser are detected but there is no near-infrared or centimeter emission. Infall (up to 2.9 \kms) 
has been identified by \citet{herpin2009} from CS data. W43-MM1 resembles a low-mass Class 0 protostar, i.e., a YSO in its main accretion phase, but scaled up in mass and luminosity. However, although it looks like a very young mid-IR quiet HMPO, W43-MM1 has already developed a hot core with temperatures higher than 200 K \citep[][]{motte2003,marseille2010}, making this object very peculiar. All these characteristics make W43-MM1 a promising massive object to study, potentially rich in water: among first 10 massive sources, from different evolutionary stages, observed by \citet{marseille2010a}, W43-MM1 is one of the three only sources (with DR21OH and IRAS18089-1732) to be detected in the \hoO~ line. 

We present here new Herschel-HIFI observations of fourteen far-IR water lines (\water, \watersept, H$_2^{18}$O), providing the most complete coverage of the water energy ladder to date, plus CS(11--10) and C$^{18}$O(9--8) data. Sections \ref{sec:observations} and \ref{sec:results} present our observations and results, whose analysis is given in Sect. \ref{sec:analysis}. Then we model the observations using a radiative transfer code in Sect. \ref{sec:model}. We estimate outflow and infall velocities, turbulent velocity, molecular abundances, and the physical structure of the source. Finally, we discuss (Sect. \ref{sec:discussion}) the results in terms of massive star formation scenarios.

\begin{table*}
\caption{Herschel/HIFI observed line transitions in W43-MM1. Frequencies are from \citet{pearson1991}. Beam and $\eta_{\textrm{mb}}$ are from \citet{roelfsema2012}. The rms is the noise at the given spectral resolution $\delta \nu$. The energy of the upper level, $E_u$, is considered to be the same for H$_2^{17}$O and H$_2^{18}$O . }             
\label{table_transitions}      
\centering                          
\begin{tabular}{lccccccccccc}        
\hline\hline                 
Water species & Frequency &  Wavelength & $E_u$    &  HIFI & Beam  &  $\eta_{\textrm{mb}}$  &  $T_{\textrm{sys}}$  &   $\delta \nu$  & t$_{int}$ & rms & obsid\\   
                      &    [GHz]     &       [$\mu$m] &      [K]  &  band &  [\arcsec]  &  & [K] & [MHz] & [s] & [mK] & \\
\hline                        
   \hoA$^a$    & 547.6764     &  547.4 &  60.5   & 1a & 38.7  & 0.754 & 80 & 0.24 & 1462 & 43 & 1342219193 \\
   \hoB    & 552.0209    &  543.1  & 61.0    &  1a & 38.0 &  0.754 & 71 & 0.24 &  158 & 46 & 1342192345 \\     
   \hoF     & 994.6751  &  301.4  &  100.6  & 4a & 21.3   &  0.741 & 288 & 0.12 & 502 & 107 & 1342194990 \\
   \hoG    & 1095.6274  &  273.8  &  248.7  &  4b & 19.2  &  0.737 & 379 & 0.24 & 1723 & 54 & 1342194806 \\
   \hoI    & 1101.6982  &  272.1  &   52.9  &  4b & 19.2 &  0.736 &  393 & 1.1 & 3566 & 17  & 1342191670,1342207372 \\
   \hoJ    & 1107.1669  &  272.1  &   52.9  &  4b & 19.1 &  0.737 &  379 & 0.24 & 1723 & 29 & 1342194806 \\
   \hoM    & 1662.4644  &  180.3  &  113.6  & 6b  & 12.8 &  0.708 & 1407 & 1.1 & 1212 & 176  & 1342192575 \\
\hline
   \hoC$^a$    & 556.9361    &  538.3  &   61.0  &  1a & 38.0  & 0.754 & 80 & 0.24 & 1462 & 43  & 1342219193 \\
   \hoD    & 752.0332    &  398.6  &  136.9  & 2b & 28.2   &  0.749 &  88 & 0.12 & 262 & 88 & 1342194565\\
   \hoE     & 987.9268 &  303.5  &   100.8 &  4a & 21.3  & 0.741 &  340 & 1.1 & 418 & 67 & 1342191616\\
   \hoH    & 1097.3651  &  273.2  &  249.4  & 4b  & 19.2  &  0.737 & 379 & 0.24 & 1723 & 54  & 1342194806 \\
   \hoK    & 1113.3430  &  269.0  &  53.4  & 4b & 19.0   &  0.736 & 393 & 1.1 & 3566 & 17 & 1342191670,1342207372\\
   \hoL    & 1661.0076  &  180.5  &  194.1  &  6b & 12.8  &  0.708 & 1407 & 1.1 & 1212 & 225 & 1342192575\\
   \hoN    & 1669.9048  &  179.5  &   114.4  &  6b  & 12.7  &  0.708 & 1407 & 1.1 & 1212 & 249  & 1342192575\\
\hline
  CS ($11-10$)   & 538.6888    &  556.5  & 155.1    & 1a & 38.0 &  0.754 & 71 & 0.24 &  158 & 46  & 1342192345 \\     
   C$^{18}$O ($9-8$)  & 987.5604 &  303.6  & 237.0 & 4a & 21.3  & 0.741 &  340 & 0.5 & 418 & 54 & 1342191616 \\
\hline                                  
\end{tabular}
\tablefoot{$^a$ This line was mapped in OTF mode. \\
}
\end{table*}

\section{Observations}
\label{sec:observations}

Fourteen water lines as well as the CS(11--10) and C$^{18}$O(9--8) lines (see Table 1) have been observed with HIFI at frequencies between 538 and 1670 GHz towards W43-MM1 in March, April, and October, 2010 (OD 293, 295, 310, 312, 333, 338, 339, and 531). The position observed corresponds to the peak of the mm continuum emission from \citet{motte2003} (RA=18:47:47.0, DEC=$-$01:54:28 J2000). The observations are part of the WISH GT-KP. 

Data were taken simultaneously in H and V polarizations using both the acousto-optical Wide-Band Spectrometer (WBS) with 1.1 MHz resolution and the digital auto-correlator or High-Resolution Spectrometer (HRS) providing higher spectral  resolution. We used the Double Beam Switch observing mode with a throw of 3'. HIFI receivers are double sideband with a sideband ratio close to unity. The frequencies, energy of the upper levels, system temperatures, integration times and {\it rms} noise level at a given spectral resolution for each of the lines are provided in Table 1. Calibration of the raw data into the $T_A$ scale was performed by the in-orbit system \citep[][]{roelfsema2012}; conversion to $T_{mb}$ was done using a beam efficiency given in Table~\ref{table_transitions}  and a forward efficiency of  0.96. The flux scale accuracy is estimated to be between 10\% for bands 1 and 2, 15\% for band 3, 4, and 20 \% in bands 6 and 7. Data calibration was performed in the Herschel Interactive Processing Environment \citep[HIPE,][]{ott2010} version 6.0. Further analysis was done within the CLASS\footnote{http://www.iram.fr/IRAMFR/GILDAS/} package.  These lines are not expected to be polarized, thus, after inspection, data from the two polarizations were averaged together. For all observations, eventual contamination from lines in the image sideband of the receiver has been checked and none was found. Because HIFI is operating in double-sideband, the measured continuum level (in the Figures and the Tables) has been divided by a factor of 2.


\section{Results}
\label{sec:results}

\begin{figure}
\centering
\includegraphics[width=\columnwidth]{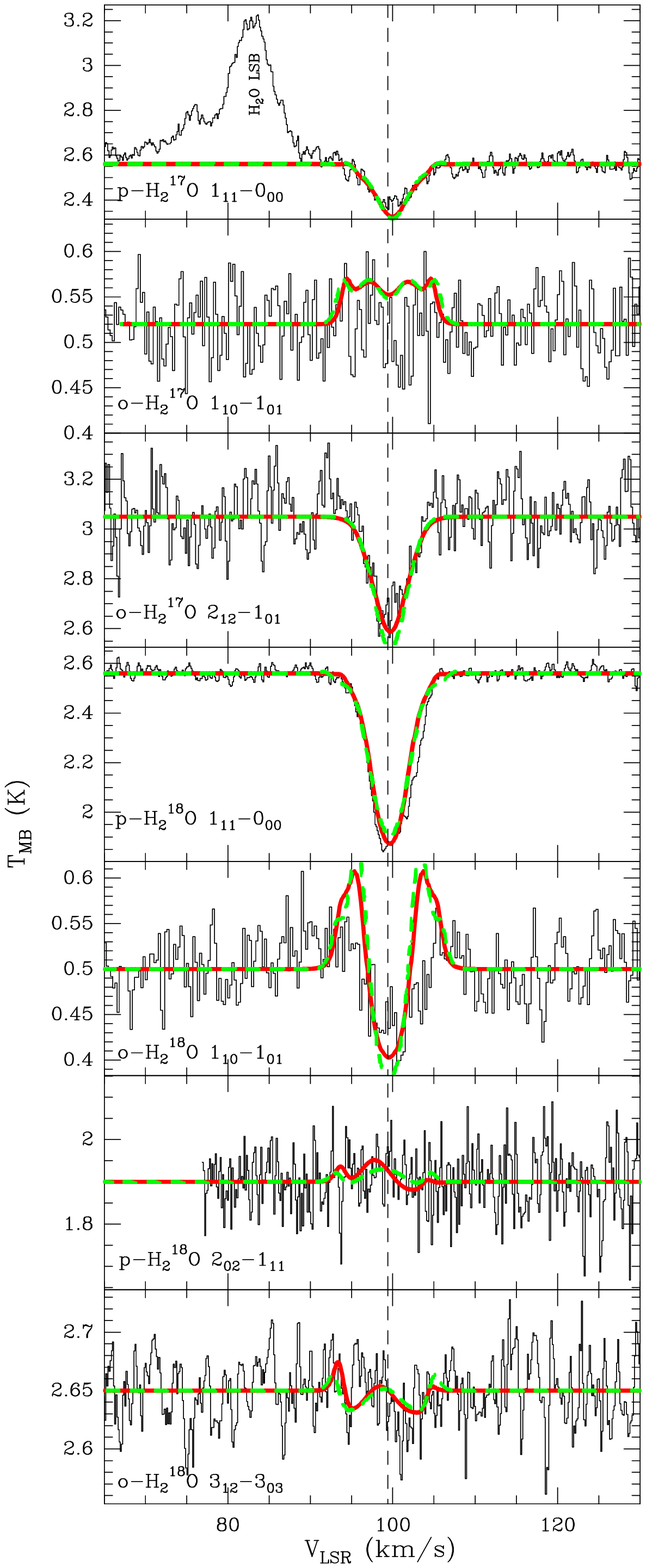}
\caption{HIFI spectra of \watersept~  and \waterhuit~ lines (in black), with continuum. The best model fits are shown as red lines over the spectra. Green dashed lines are the model fits with constant V$_{tur}$= 2.5 \kms and V$_{infall}$= 0.0 \kms. Vertical
dotted lines indicate the \vlsr~ at 99.4 \kms. The spectra have been smoothed to 0.2 \kms, and the continuum divided by a factor of 2.}
\label{fig:lines_iso}
\end{figure}

\begin{figure}
\centering
\includegraphics[width=\columnwidth]{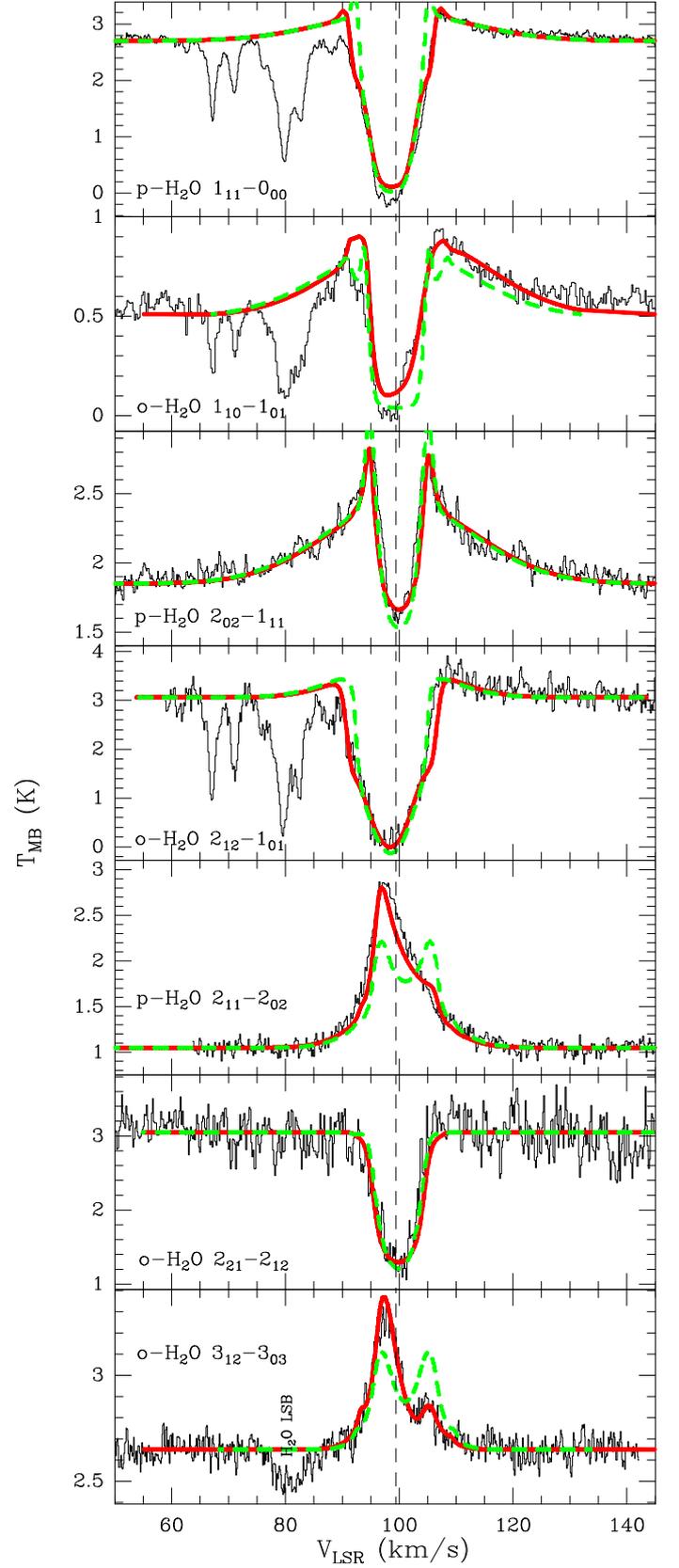}
\caption{HIFI spectra of \water~ lines (in black), with continuum. The best model fits are shown as red lines over the spectra. Green dashed lines are the model fits with constant V$_{tur}$= 2.5 \kms and V$_{infall}$= 0.0 \kms. Vertical dotted lines indicate the \vlsr~ at 99.4 \kms. The spectra have been smoothed to 0.2 \kms, and the continuum divided by a factor of 2.}
\label{fig:lines_H2O}
\end{figure}

\begin{figure}
\centering
\includegraphics[width=\columnwidth]{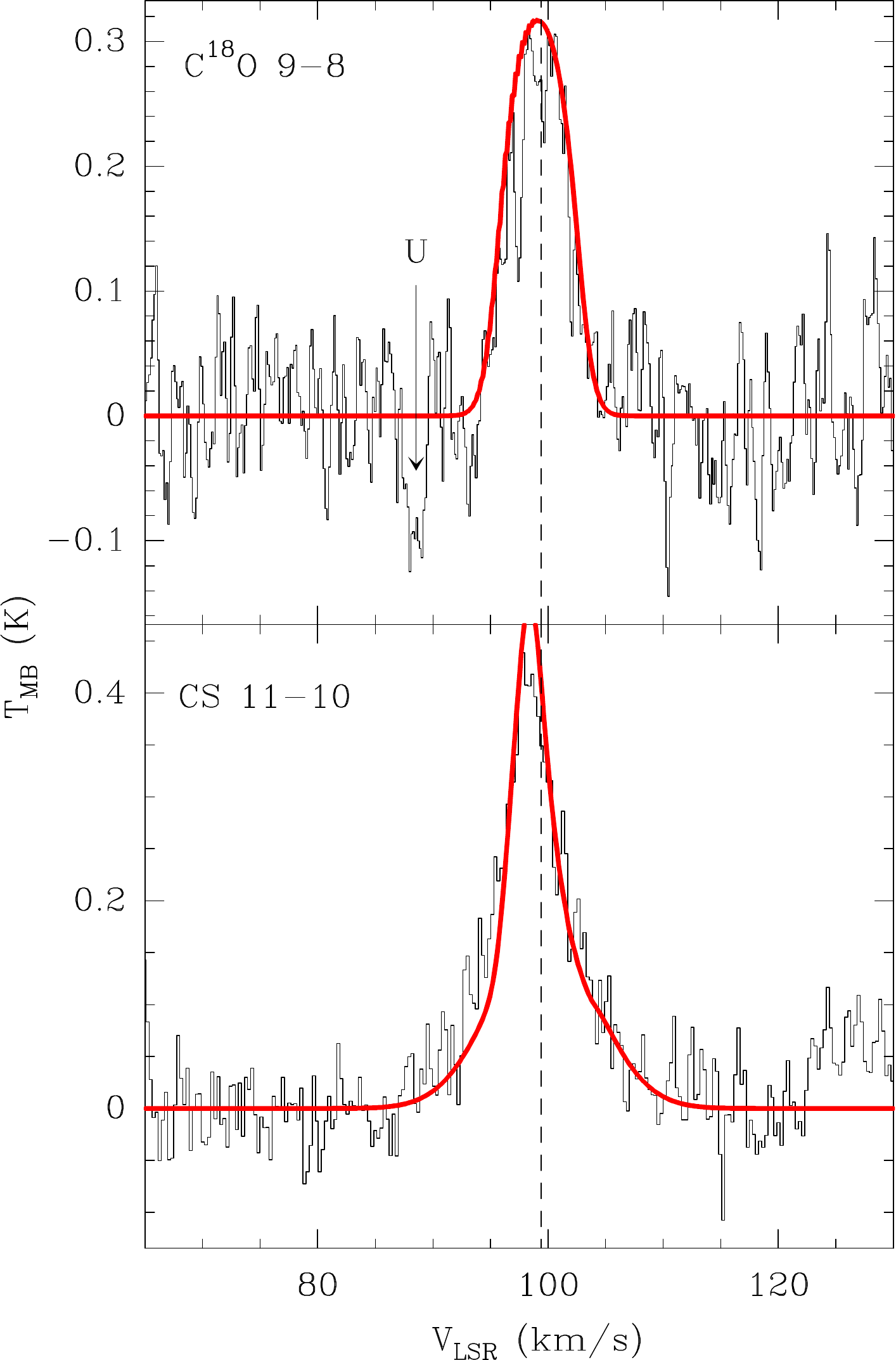}
\caption{Continuum subtracted HIFI spectra of the C$^{18}$O 9-8 and CS 11-10 lines (in black). The model fits are shown as red lines over the spectra. Vertical
dotted lines indicate the \vlsr~ at 99.4 \kms.The spectra have been smoothed to 0.2 \kms. The unidentified line in absorption seen at 88.6 \kms~ is indicated.}
\label{fig:lines_c}
\end{figure}

The spectra including continuum emission are shown in Figures \ref{fig:lines_iso} and \ref{fig:lines_H2O} for the rare isotopologues (\watersept, H$_2^{18}$O) and for \water~ respectively. In addition, Fig.\ref{fig:lines_c} displays the CS ($11 - 10$) and  C$^{18}$O ($9-8$)  spectra. We show the HRS spectra, except for the \hoI, \hoM, \hoE, \hoK, \hoL, and \hoN lines where WBS spectra were used since the velocity range covered by the HRS was not sufficient.

We derive the peak main beam temperatures and half power line-widths for the different line components from multi-component Gaussian fits, made with the CLASS software (line parameters are given in Table~\ref{line-W43MM1}). All lines associated with the source have V$_{lsr}\simeq 95-100$ \kms. Since W43-MM1 is located  in a very crowded area of the galactic plane (b$\sim -0.1^{\circ}$), foreground clouds contribute to the spectra through water absorptions at  V$_{lsr}$ shifted with respect to W43 velocity in the \hoC, \hoK~ and \hoN~ lines spectra. These will be discussed in Appendix \ref{sec:absorption}.

As in \citet{johnstone2010}, \citet{kristensen2010}, and \citet{chavarria2010}, we adopt the following terminology for the different velocity components: broad (FWHM$\simeq$20-35 \kms),  medium (FWHM$\simeq$5-10 \kms), and narrow ($\sim$3 \kms). 

\subsection{C-bearing species}

Although the main goal of WISH is to observe H$_2$O, the HIFI bands included the C$^{18}$O (9-8) and CS (11-10) lines  as well. They are seen in emission without any broad component contribution. While the source velocity derived by \citet{motte2003} from H$^{13}$CO$^+$ observations is 98.8 \kms, the best Gaussian fit for the observed C$^{18}$O ($9-8$) line emission is centered at 99.4 \kms~ (with a FWHM of 5.6 \kms), which is consistent taking into account the uncertainties. 
The CS 11-10 line profile is best fit by two Gaussians of 2.9 and 10.2 \kms width, centered at 98.4 and 98.8 \kms, respectively, but the CS line profile is asymmetric, possibly revealing infall, blue-shifting the emission peak. In the following, we adopt  V$_{lsr}=99.4$ \kms as the source velocity. 

Note that an unidentified line is detected in absorption in the C$^{18}$O (9-8) spectra at 88.6 \kms, i.e., 987.594$/$972.747 GHz in USB$/$LSB).

\subsection{\water~ lines}

Most of the \water~ lines can be described as the sum of a medium (FWHM$\simeq$5-10 \kms) and a broad (up to 35 \kms) velocity components. The broad one is likely the signature of an outflow. The observed dependence of the outflow velocity extent with $E_{up}$ is probably due to sensitivity effects (the outflow being more difficult to detect with weaker line intensity). The medium component is centered at 99.4-99.9 \kms~ (fitting uncertainty) while the broad one is redshifted by a few \kms from the source velocity. Main water line profiles are more complex than for the rare isotopologues. Except for the \hoL, \hoD, and \hoH~ lines, none of the spectra exhibit pure absorption or emission line profiles. 

For low-mass objects, \citet{kristensen2010} interpret the narrow emission as the signature of the passively heated envelope while the medium component might be due to shocks, but on spatial scales smaller than the molecular jets; the broad emission  arises in the extended molecular outflow. 

The para and ortho ground-state lines are deeply absorbed but their profiles exhibit a broad outflow component too. The \hoC~ and \hoK~ line profiles are fully absorbed as is more or less the \hoN~  line while the  peak depth of the \hoL~ line is 1.1 K ($\sim 65 \%$ of the continuum level, see Table \ref{line-W43MM1}). The \hoE~ line profile is also dominated by an absorption dip at 1.6 K ($\sim 11 \%$ of the continuum), but reveals a strong (up to 1K) and broad outflow (21 \kms) component. This line is actually the most sensitive to the outflow. 

Except for the \hoL~ line with $E_{up}$ at 194 K which is seen in absorption, the emission lines involving energy levels above 130 K reveal an infall signature characterized by an asymmetrical profile with a stronger blue component compared to the red one (see for example the \hoD~ spectra).

\subsection{Rare water isotopologues}

Among all observed \watersept~ and \waterhuit~ lines, the \hoB~ and \hoF~ lines are not detected (assuming a line width of 6 \kms, upper limits are 280 and 650 mK \kms, respectively). The  \hoG~ line is tentatively detected at 2 $\sigma$ level.  

The observed profiles of lines connecting with a ground-state ortho or para level are dominated by absorption from the cold outer envelope. 
This is a bit surprising as one would have expected at least the \hoM~ to be in emission because of the high energy level involved. All these absorptions are centered at $99.6\pm0.3$ \kms and can be fit by a single Gaussian with a velocity width ranging from of 4.7 to 6.4 \kms, corresponding to our "medium" component. 
No broad component is observed. Line parameters are given in Table \ref{table_param}. 


\begin{table*}
  \caption{Observed line emission parameters for the detected lines. $\varv_{LSR}$ is the line peak velocity. $T_\mathrm{mb}$ is the intensity of the main peak (i.e. of the strongest emission$/$absorption). $T_{cont}$  is the (corrected) continuum level measured from the corresponding WBS spectra. $\Delta\varv$ are the velocity full width at half-maximum (FWHM) of the narrow, medium and broad components. }\label{line-W43MM1}
\begin{center}
\label{table_param}      
\begin{tabular}{lccccccccc} \hline \hline
{\bf Line}  & $\varv_{LSR}$ & T$_\mathrm{mb}$ & T$_{cont}$ & $\Delta\varv_{nar}$ & $\Delta\varv_{med}$ & $\Delta\varv_{br}$ & $\tau$ & F & \Lsun \\ 
 & [km$/$s] & [K] & [K] & [km$/$s] & [km$/$s] & [km$/$s] & & [$10^{-21}$ W.cm$^{-2}$] &  \\ 
\hline                        
   \hoA     & 100.8 $\pm$0.5 &  0.40$^{b}$ $\pm$0.05 & 0.50$\pm$0.05 & - & 6$\pm$1 & -  & 0.20 $\pm$ 0.03 & & \\
   \hoG$^{a}$   & 101.2$\pm$0.2 & 2.6$^{b}$$\pm$0.4 & 2.6$\pm$0.4 & - & - &  - &  0.02 $\pm$ 0.01 & & \\
   \hoI     & 99.8$\pm$0.3 & 1.8$^{b}$$\pm$0.3 & 2.6$\pm$0.4 & - & 5.8$\pm$0.3 & - & 0.33 $\pm$ 0.07  & & \\
   \hoJ    & 99.6$\pm$0.1 &  2.4$^{b}$$\pm$0.4 & 2.6 $\pm$0.4& - & 6.4$\pm$0.2 & -  & 0.08 $\pm$ 0.02 & & \\
   \hoM   & 98.6$\pm$0.2 &  2.6$^{b}$$\pm$0.5 & 3.1$\pm$0.6 & - & 4.7$\pm$0.3 & -  & 0.14  $\pm$ 0.05 & & \\
\hline
   \hoC     & 99.3$\pm$0.2 & 0.0$^{b}\pm0.05$ & 0.50$\pm$0.05 & - & 7.5$\pm$0.2 & 35.5$\pm$0.2 & $>6$  & 6.0 $\pm$ 0.6  & 0.057 $\pm$ 0.006 \\
   \hoD      & 97.4$\pm$0.1 & 1.8$\pm$0.2 & 1.0$\pm$0.1 & - & 7.5$\pm$0.2 & 18.5$\pm$0.4 &  & 92 $\pm$ 9 & 0.87 $\pm$ 0.09 \\
   \hoE      & 94.8$\pm$0.3 & 1.6$^{b}\pm$0.2 & 1.8$\pm$0.3 & - & - & 21.0$\pm$0.5 & 0.14 $\pm$ 0.01 & 34 $\pm$ 5 & 0.32 $\pm$ 0.05 \\
   \hoH   & 98.0$\pm$0.2 & 0.6$\pm$0.1 & 2.6$\pm$0.4 & - & 5.9$\pm$0.2 & - &  & 7 $\pm$ 1 &  0.064 $\pm$ 0.009\\
   \hoK    & 97.8$\pm$0.2 & 0.0$^{b}\pm$0.02  & 2.6$\pm$0.4 & - & 10.3$\pm$0.2 & 30.5$\pm$0.6 &   $>6$  & 8 $\pm$ 1 & 0.07 $\pm$ 0.01  \\
   \hoL    & 101.1$\pm$0.2 & 1.1$^{b}\pm$0.3 & 3.0$\pm$0.6 & - & 6.8$\pm$0.2 & -  & 1.0  $\pm$ 0.1 &  &\\
   \hoN    & 97.9$\pm$0.2 & 0.0$^{b}\pm$0.3 & 3.0$\pm$0.6 & - & 11.8$\pm$0.4 & 20.5$\pm$0.6 & 5.72   $\pm$ 0.02  & 8 $\pm$ 2 & 0.08 $\pm$ 0.02 \\
\hline
  CS $11-10$   & 98.3$\pm$0.3 & 0.42$\pm$0.04 & 0.50$\pm$0.05 & 2.9$\pm$0.9 & 10$\pm$1 & - & & \\     
   C$^{18}$O $9-8$  & 99.4$\pm$0.3 & 0.30$\pm$0.05 & 1.8$\pm$0.3 & - & 5.6$\pm$0.3 & -  & & \\
  \hline 
\end{tabular}
\end{center}
\tablefoot{$^a$ tentative detection.
$^b$ Line in absorption. Temperature for the absorption dip with continuum \\}
\end{table*}

\section{Analysis}
\label{sec:analysis}

\subsection{Line Asymmetries}

Outflows, infall, and rotation can produce very specific line profiles with characteristic signatures \citep[see][and references therein]{fuller2005}. Outflow and rotation give rise to both red and blue asymmetric lines while the profiles of optically thick lines from infalling material have stronger blue-shifted emission than the redshifted emission. However, in some circumstances, outflow or rotation could also produce a blue asymmetric line profile along a particular line of sight to the source. 


Compared to the C$^{18}$O ($9-8$) optically thin line ($\varv=99.4~ $\kms), several water lines profiles show clear asymmetry.  
Focusing our analysis on emission lines, strong asymmetries are detected in the \hoD~ and \hoH~ lines. Both profiles clearly exhibit a stronger blue component, presumably indicating the presence of infall. Actually, strong asymmetries have previously been measured in the CS line profiles (involving lower energy levels) towards this source by \citet{herpin2009}, and the CS 11-10 line observed here exhibits an infall signature too.

The absorption lines (including the rare isotopologues) do not show clear asymmetry, but as indicated by the absorption peak velocities (see Table \ref{line-W43MM1}), several of them are indeed redshifted (relative to the source velocity of 99.4 \kms), as expected in case of infall.

The \hoE~ line profile, with a very sightly stronger blue peak than the red one, exhibits a strong self-absorption dip at the source velocity. This almost symmetric double-horn profile might be produced by the outflow \citep[][]{fuller2005}. Actually, the absorption is seen against both the outflow and continuum emission. Hence, as shown by Kristensen et al. (in press), the outflow is also embedded (i.e., the absorption layer is in front of both the emitting layers).  

\subsection{Opacities}

For the lines in absorption, we estimate the opacities at the maximum of absorption from the line-to-continuum ratio in  Table \ref{line-W43MM1} using the formula
\begin{equation}
\tau = -ln (\frac{T_{mb}}{T_{cont}})
\end{equation} 
and assuming that the continuum is completely covered by the absorbing layer.

The \waterhuit~ lines are somewhat optically thick, with $\tau \approx$ 0.2--0.3, if we exclude the tentatively detected \hoG~ line. In contrast, the \watersept~ lines are close to optically thin, with $\tau \approx 0.1$.

Of the \water~ lines detected in absorption, all but the \hoE~ line are highly optically thick. Note that the optical depth of this line may be (much) higher if not just continuum but also line emission is absorbed (see van der Tak et al. in preparation). The \hoC, \hoK~ and \hoN~ line profiles show absorption nearly down to the zero-temperature level, leading to opacities larger than 3 in the coolest shells. All these absorptions are due to cold material in which the protostar is embedded.   

\section{Modeling}
\label{sec:model}
The preceding sections have demonstrated how the line profiles can be decomposed into various dynamical components empirically and how their asymmetry can signal the presence of infall. In this section,
we model the full line profiles in a single spherically symmetric model with different kinematical components due to turbulence, infall, and outflow.

\subsection{Method}
\label{sec:method}

\begin{figure}
\centering
\includegraphics[width=9.5cm]{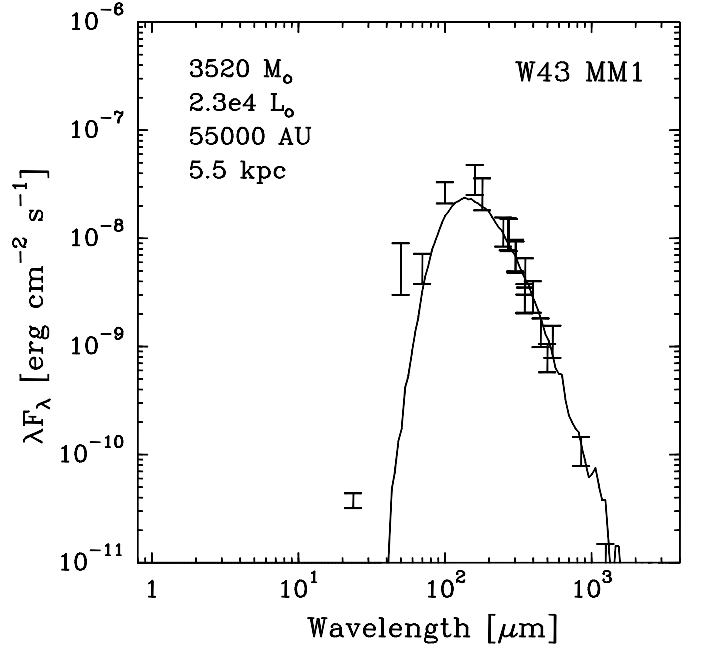}
\caption{Spectral energy distribution for W43-MM1 calculated using the Whitney-Robitaille model for the parameters in Table \ref{table_parameters}. Error bars are from the references cited in Sect. \ref{sec:method}.}
\label{fig:sed}
\end{figure}

We use the 2-dimensional Whitney-Robitaille radiative transfer code \citep[hereafter WR,][]{whitney2003,robitaille2006,robitaille2007} to derive the envelope temperature and density profiles. The physical model consists of a spherical envelope with no disk and no cavity to emulate a 1-dimensional model. The density decreases exponentially with radius and it follows the relation:
\begin{equation}
n(r)=n_o \left(\frac{r}{r_o}\right)^p,
\end{equation}
where $n_o= 5.77\times10^6$ cm$^{-3}$ is the number density at a radius $r_o=1\times10^4$ AU for the derived envelope mass. We assume a $p$ value of $-1.5$  (model parameters are listed in Table \ref{table_parameters}). Dust opacities (with thin ice coated grains) are taken from \citet{ossenkopf1994}. The envelope temperature and mass are constrained by comparing the WR derived spectral energy distribution (SED) with existing sub-millimeter continuum fluxes (seeFig. \ref{fig:sed}) from our Herschel observations and \citet{bally2010}, CSO/SHARC and IRAM/MAMBO observations from \citet{motte2003}, JCMT/SCUBA and SPITZER-MIPS archive data (see Fig. \ref{fig:sed}), and KAO observations \citep[][]{lester1985}. Observed peak fluxes were normalized to the source size (55000 AU $\simeq$ 10" at 5.5 kpc) derived from 1.2 mm continuum at 3 sigma over noise level \citep[][]{motte2003}. 



\begin{figure}
\centering
\includegraphics[width=9.cm]{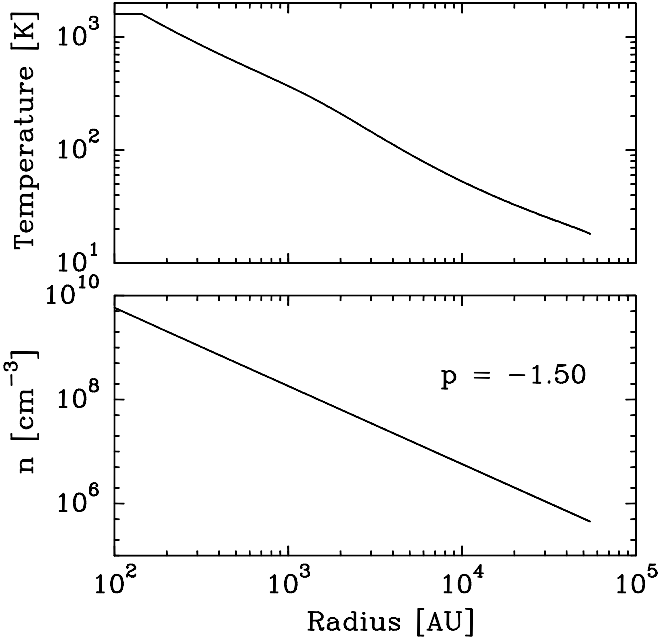}
\caption{Density and temperature profile derived from SED modeling.}
\label{fig:TN}
\end{figure}

Line emission is modeled with the 1D-radiative transfer code RATRAN  \citep[][]{hogerheijde2000} using the WR derived envelope temperature and density profiles (see Fig.\ref{fig:TN}), following the method described in \citet{marseille2008} and \citet{chavarria2010}. The H$_2$O collisional rate coefficients are from \citet{faure2007}.

Our model of W43-MM1 has two components: an outflow and the proto-stellar envelope.  
The outflow parameters (intensity, velocity width) are derived from the Gaussian fitting of the emission lines, with FWHM between 10.2 and 35.5 \kms depending on the line. We have used RADEX \citep[][]{vandertak2007} to estimate the H$_2$O column density and opacity in the outflow. To retrieve the observed intensities for the outflow component we adopt $T_{kin}$ = 100 K, $n(H_2) = 10^7$ cm$^{-3}$ (following the density profile shown on Fig. \ref{fig:TN}), and a water column density of $10^{14}$ cm$^{-2}$. Because RATRAN only enables to include the outflow as an additional component (with T, opacity, and width) not part of the radiative transfer process, the outflow component is hence assumed to be an unrelated component, both spatially and in velocity. The result is directly incorporated into the ray-tracing part of the RATRAN code \citep[see][]{hogerheijde2000}. As a consequence, we stress that the absorption in the modeled line profile is very likely overestimated, i.e., the cold layer in size or temperature.

The envelope contribution is parametrized with three input variables: water abundance ($\chi_{H_2O}$), turbulent velocity (V$_{tur}$), and infall velocity (V$_{infall}$). The width of the line is adjusted by varying V$_{tur}$. The line asymmetry is reproduced by the infall velocity V$_{infall}$. The line intensity is best fitted by adjusting a combination of the abundance, V$_{tur}$, and V$_{outflow}$ parameters. We adopt the following standard abundance ratios for all the lines: 450 for \water$/$\waterhuit, 4.5 for \waterhuit$/$\watersept~ \citep[][]{wilson1994, thomas2008}, and 3 for ortho$/$para-H$_2$O. The models assume a jump in the abundance in the inner envelope at 100 K  (see Sect. \ref{abun_struc}). Table \ref{table_physical} gives the parameters used in the models. 

To fit the observed lines we first modeled the rare isotopologues (\watersept, then \waterhuit) lines and the \water~ line involving the highest energy level (\hoH; optically thin), using the same abundances for all lines (\water~ abundances are derived from \watersept~ and \waterhuit~ values times the isotopic abundance ratios). Once we are able to reproduce the main features of the profiles by minimizing residuals in a grid of values, we model the rest of the lines using the same parameters, including the outflow component when justified.

\begin{table}
\caption{Parameters used in Whitney-Robitaille  continuum model and derived physical quantities.}             
\label{table_parameters}      
\begin{tabular}{l r | l r}        
\hline\hline                 
 \multicolumn{2}{c}{Parameter used}  & \multicolumn{2}{| c}{Derived parameter} \\    
\hline                        
Inner radius (AU)$^{a}$                 & 100   &  Mass (\msun) & 3520        \\
Outer radius (AU)$^{a}$                 & 55000   & $T_{out}$ (K) & 19      \\
Distance (kpc)			& 5.5 & $T_{in}$ (K) & 1377  \\	
Luminosity (\lsun)               & 2.3$\times$ 10$^4$ & &           \\
Density exponent 	& -1.5 & & \\
\hline                                   
\end{tabular}
\tablefoot{$^a$ \citet{motte2003} \\}
\end{table}

We simultaneously use two different criteria to quantify the quality of our model: 
\begin{itemize}
  \item we quantify the error $\varepsilon$ relative to the intensity $T$ and width $\Delta v$ of the line
\begin{equation}
\varepsilon = \frac{1}{2} \Bigg{|}\frac{T_{peak,mod}-T_{peak,obs}}{T_{peak,obs}}\Bigg{|}+\frac{1}{2} \Bigg{|}\frac{\Delta v_{mod}-\Delta v_{obs}}{\Delta v_{obs}}\Bigg{|}
\end{equation}
 \item the similarity between the observed and the modeled line profiles is quantified by comparing for each channel the observed ($T_i$) and modeled intensities (${T_i}^{\star}$) (above the 3$\sigma$ detection limit). We define a detection function $D_i$ and a correlation function $C_i$ as follows 
\begin{equation}
D_i =
\left \{ 
\begin{array}{rl}
1 & {\textrm{if}}~ T_i > 3\sigma ,\\ 
0 & {\textrm{or}}
\end{array} 
\right.
and~
%
C_i =
\left \{ 
\begin{array}{rl}
1 & {\textrm{if}}~ |T_i -{T_i}^{\star}| <3\sigma ,\\ 
0 & {\textrm{or}}
\end{array} 
\right.
\end{equation}
The computed parameter is called the profile similarity factor $\Sigma$
\begin{equation}
\label{ }
\Sigma=\frac{\sum_i C_iD_i}{\sum_iD_i}
\end{equation}
\end{itemize}
Values of $\varepsilon$ and $\Sigma$ are respectively minimized and maximized by varying the parameters in Table \ref{table_physical}. For consistency purpose with other works, we have also checked the classical $\chi^2$ values.

\subsection{Velocity structure}
\label{velocity}
In order to constrain the source dynamics, we try two different approaches: 1) a model with constant turbulent  velocity and zero infall (V$_{tur}$= 2.5 \kms and V$_{infall}$= 0.0 \kms) for all lines and all radii; 2) a model in which the velocity parameters vary with radius. Inspection of the line profiles (see Sect. \ref{sec:results})  shows that the width of the velocity components is not the same for all lines, nor is the infall contribution, so we do not expect a model with equal velocity parameters for all lines to fit the data well. Indeed, for most of the lines, the model 1 (see Figs. \ref{fig:lines_iso} and \ref{fig:lines_H2O}) fails to reproduce the global shape of the profiles, especially those of \water.

The best model has turbulence and infall both varying with the radius (see Figs. \ref{fig:lines_iso} and \ref{fig:lines_H2O}), as already seen by \citet{caselli1995} for the turbulence. We first tried to apply a power-law variation for both parameters but without any success. We also unsuccessfully tried a "two-steps" turbulence ($V_{tur}$= 2.2 \kms~ up to a certain radius, i.e., around $10^4$ AU, and 3.5 \kms~ beyond). A "multiple steps profile" (Fig.\ref{fig:turb}) for the turbulence proved to be a much better approach to fit all line profiles with the same model. The infall is also best constrained by a step function given in Fig.\ref{fig:turb}. Each shell of the envelope (at radius $R$) is hence characterized by a temperature, a density, a turbulent, and an infall velocity as shown in Fig.\ref{fig:TN} and \ref{fig:turb}. We have estimated the uncertainties of these velocities by making the velocity varying around the "best value " at each radius.

Line profiles for all species are well reproduced by our model (Figs. \ref{fig:lines_iso}, \ref{fig:lines_H2O} and \ref{fig:lines_c}), with $\Sigma$ similarity coefficient larger than 90\%, and minimized $\varepsilon$ and $\chi^2$ for most lines (\hoA, \hoD, and \hoC~ have $\Sigma \simeq 75$\%), even with an outflow contribution fit with a simple Gaussian model. The model predictions are consistent with the upper limits for the non-detected lines \hoB~ and \hoF, and the tentatively detected \hoG~ line. The CO and CS lines are well fit by the model derived from the other lines. 
All modeled lines are centered on V$_{LSR}$ equal to 99.4 \kms, but the outflow component central velocity varies somewhat from line to line by $\pm0.5-1.0$ \kms. In addition, we apply our model to the \waterhuit~ $3_{13}-2_{20}$ line at 203.3916 GHz ($E_{up}$=204 K) observed with the IRAM-30m telescope ($\Delta$V = 3.8 \kms) by \citet{marseille2010a}. The line profile is perfectly reproduced.

\subsection{Infall and accretion rate}

Actually, outflow \citep[e.g.,][]{walker1988} and rotation \citep[][]{zhou1995, fuller2005} can also give rise to blue asymmetric lines along a particular line of sight to the source. But considering that blue asymmetry is observed in several water transitions \citep[see also][for CS]{herpin2009}, and that there is no evidence of contradictory information in the form of red asymmetric profiles, the infall explanation is the most likely. Nevertheless, only mapping (and position-velocity diagram study) with a high angular resolution (of a few 0.1 arcsec) could help to separate the effects of rotation and infall. Indeed, rotation might produce a symmetry reversal in outer part in direction perpendicular to the rotational axis where the optical thickness is less \citep[][]{zhou1995, yamada2009}. 

From the infall velocity $V_{infall}$ we estimate the mass accretion rate $\dot{M}_{acc}$ applying the method described by \citet{beltran2006}. As a first rough estimate we assume that the gas is undergoing spherically symmetric free-fall collapse onto the central object. For any radius $R$, the mass accretion rate is given by:
\begin{equation}
\label{infallrate}
\dot{M}_{acc} =4\pi R^2 m n~ V_{infall},
\end{equation}
where $m$ is the mean molecular mass, $n$ the gas volume density, and $V_{infall}$ the infall velocity at $R$ as shown on Fig. \ref{fig:turb}.

The maximum infall velocity of 2.9 \kms~ is mainly derived from the \hoD~ and \hoH~ line profiles. From the model (Fig. \ref{fig:turb} and Sect. \ref{velocity}) we can estimate for that infall velocity the corresponding radius from the protostar to $5.7-7.3 \times 10^{17}$ cm  (38100-48800 AU). The inferred mass accretion rate for this range of radii is then 3.5 to 4.0 $\times$10$^{-2}$~M$_{\odot} $yr$^{-1}$, slightly lower by a factor 2 than what is derived by  \citet{herpin2009} from CS observations. At such radii, we are actually likely measuring the accretion from the envelope toward the disk, if there is a disk. However, according to our model, the infall velocity is varying with radius, i.e., the mass accretion rate is not constant with radius. Inside of 26800 AU, the infall velocity is only 0.4 \kms, which leads to a mass accretion rate between 4.0 $\times$10$^{-4}$ and 5.0 $\times$10$^{-3}$~M$_{\odot} $yr$^{-1}$.

\subsection{Abundance structure}
\label{abun_struc}

Modeling the whole set of observed lines allows us to determine the abundance structure. Among the lines mostly excited in the inner envelope, only the higher-$J$ \waterhuit~ (\hoG~ and \hoO) and the \hoH~ are optically thin enough to probe the inner part of the envelope. An abundance jump at 100 K is necessary to reproduce these lines while the other lines are not sensitive to it. The derived \water~ abundance relative to H$_2$ is  1.4 $\times 10^{-4}$ in the warm part where $T>$100 K while it is 8.0 $\times 10^{-8}$ in the outer part where $T<$100 K, in agreement with the values found by \citet{marseille2010a} from \waterhuit~ observations. No deviation from the standard o$/$p ratio of 3 at high-temperature ($\geq$100 K) is found. Hence, the one-dimensional approach seems acceptable as the shape of all line profiles is well recovered.
 
From the integrated fluxes given in Table \ref{line-W43MM1} (measured for the lines at least partly in emission), one can derive the water luminosities. We then estimate, assuming isotropic radiation and at a distance of 5.5 kpc, that the minimum  total {\em HIFI} water luminosity is 1.5 \lsun (sum of all individual observed luminosities). The true water emission from the inner part could be much greater but the cool envelope absorbs much of the emission. Moreover, from the modeling we estimate the total water {\bf (radial)} column density in the envelope to be 1.56 $10^{21}$ cm$^{-2}$, corresponding to 0.11 \msun~ (M$_{H_2^{18}O}=2.4\times10^{-4}$ \msun ~and M$_{H_2^{17}O}=5.6\times10^{-5}$ \msun). More than 97\% of this mass is in the inner part. The CS and C$^{18}$O column densities are respectively $7.3\times10^{15}$ and $8.2\times10^{17}$ cm$^{-2}$, leading to masses of $3.2\times10^{-4}$ (CS) and  $1.0\times10^{-2}$ (C$^{18}$O, 5.0 for CO) \msun. Hence the mass of oxygen trapped in H$_2$O and CO is around 3 \msun, 96\% locked in CO and 4\% in H$_2$O.

\begin{table}
\caption{Physical quantities derived from the RATRAN model.}             
\label{table_physical}      
\begin{tabular}{l r }        
\hline\hline                 
Parameter  &  \\    
\hline                        
\agua                                   &  8.0 ($\pm$1.0) $\times10^{-8}$  \\
Post-jump \agua                 &   1.4 ($\pm$0.4) $\times10^{-4}$ \\
\ratioop & 3 $\pm$0.2 \\
\ratiosept & 4.5 \\
\ratiohuit &Ê450 \\
\vtur~(\kms)                      & 2.2-3.5                          \\
V$_{outflow}$~(\kms)                     & 10.2-35.5                    \\
V$_{\textrm{\tiny{infall,max}}}$~(\kms)                     & -2.9                    \\
\vlsr ~(\kms)                      & 99.4          \\
\hline                                   
\end{tabular}
\end{table}

\subsection{Uncertainties and robustness of the parameters}

Uncertainties in our calculations result mainly from three sources:
\begin{itemize}
  \item the intensity calibration of the observed lines (see Sect. \ref{sec:observations}).
  \item The physical limits of our model whose radiative transfer computations are based on a physical description of the source derived from a 1D SED modeling. 
  \item The parameter sensitivity of the line radiative transfer model. 
\end{itemize}

We have rerun the models first increasing, then decreasing the observed line intensities by the uncertainties given in Section \ref{sec:observations}. We found  differences in the estimated abundances between 10 and 30\% depending on the line. To test the influence of the physical model, we decreased the envelope mass by 30\% (which gives an SED just below the error limits) and recalculate the water abundances. We found differences up to 75\%. In addition, considering that the error on the exponent of the density profile is 0.1 (see Sect. \ref{sec:method}), we have tested the robustness of our results when varying the density profile within that range: we found that the abundances are impacted by 30-40\%. Moreover, the uncertainty for the distance to the source \citep[we adopt 5.5 kpc, but][place the distance at 5.9$^{+1.2}_{-0.7}$ kpc]{nguyen2011} translates into an uncertainty on the physical model: 10\% for the distance leads to 15 \% for the mass hence 30 \%  in the abundance. This of course impacts our results. In addition, we checked that varying the best fit parameters from Table \ref{table_physical} by 5 \% has no influence on the derived abundances. Variation of the turbulent velocity within the error bars (and varying the radius by 15\%) shown on Fig.\ref{fig:turb} does not affect the results. Actually this turbulent velocity, as explained in Sect. \ref{velocity}, is well constrained thanks to the high number of observed lines, efficiently probing different radius. The determination of the infall velocity versus the radius is more uncertain because the infall signature is not visible for the quite completely absorbed lines. 

Unfortunately, we are not able to determine if these uncertainties are independent of each other. Hence, only the uncertainties coming from the HIFI intensity calibration provide us with a realistic quantitative estimate of the impact on our calculations. In a forthcoming study, we plan to investigate how a 2-dimensional model incorporating disk and cavity \citep[e.g.,][]{hosokawa2010} would impact the results presented here. The fact that the spherical model provides such a good fit argues against large changes. Nevertheless the modeling results presented here are the best we got, but it is obvious that the model is very likely not the only possible one. However, we stress that by varying parameters, the derived numbers change, but not our main conclusion that the turbulence and the accretion rate vary with radius.

\section{Discussion}
\label{sec:discussion}

\subsection{Comparison with previous work}

These observations towards W43-MM1 of a large set of water lines, together with CS and C$^{18}$O,  reveal that, as for low- and intermediate-mass objects \citep[][] {kristensen2010,yildiz2010, johnstone2010}, one finds a broad outflow ($> 20$ \kms) and a medium (5-10 \kms) velocity component. On the opposite, the narrow ($<5$ \kms) component is not clearly detected, even if it might be present in the CS line profile. They show two features not seen in lower mass types of objects. First, as noted by \citet{chavarria2010} in W3IRS5, the emission of the rare isotopologues mainly comes from the medium component in contrast with the low-mass objects where the broad outflow is the main origin. Second, no narrow component is seen in the water lines, maybe because of opacity effects. The narrow component emission comes very likely from the hot core neighborhood (i.e., the passively heated inner envelope) and observations of higher-energy level water lines are needed to probe more deeply this region. The broad component originates in the molecular outflow similar to what is observed in low-mass objects while the medium velocity component is likely due to combination of turbulence and infall. The main difference with W3IRS5 is the presence of infall and no sign of expansion. 

According to our model, the outer water abundance derived in W43-MM1 ($8 \times 10^{-8}$) is several times higher than what is found in the mid-IR bright HMPO W3IRS5 \citep[$\chi_{out}=1.8 \times 10^{-8}$,][]{chavarria2010} or in the HMC G31.41 \citep[$\chi_{out}=10^{-8}$][]{marseille2010}. It is also an order of magnitude larger than in low-mass objects \citep[$\chi_{out}=10^{-8}$,][]{kristensen2010}. In contrast, although observing lines with higher upper energy levels could help to improve the estimate, the inner abundance in W43-MM1 ($1.4 \times 10^{-4}$) can be constrained to a value similar to what is found in other high-mass sources \citep[e.g. $\chi_{in}=1.0-2.0 \times 10^{-4}$,][]{vandertak2006,chavarria2010}, i.e. , most oxygen contained in water.

\begin{figure}
\centering
\includegraphics[width=9cm, angle=0]{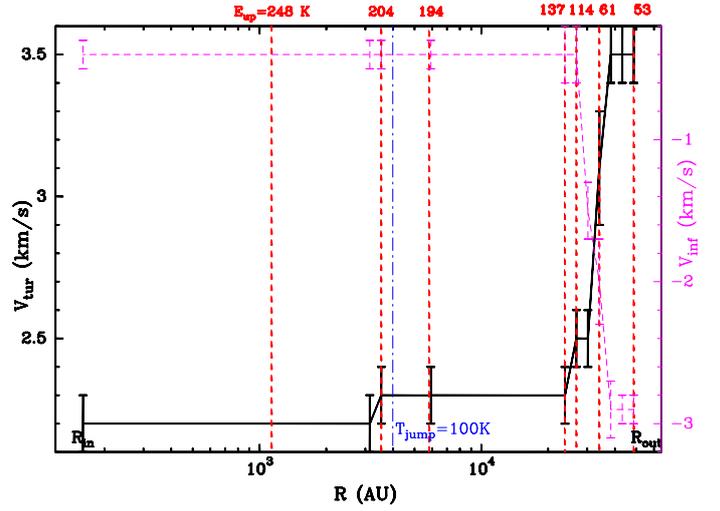}
\caption{Variation of the turbulent and infall velocities, respectively $V_{turb}$ and $V_{inf}$, with the radius $R$ (AU) to the central object. Overplotted are the radii corresponding to the temperature jump (dashed) in the model and to the radius where lines of different energies are mainly coming from with upper energy levels as dashed lines.}
\label{fig:turb}
\end{figure}

\subsection{W43-MM1 and the high-mass star formation process}

This work shows that the observed molecular emission is dominated by turbulent velocities ($v_{tur}> 2.2$ \kms) greater than the speed of sound ($a_S\simeq0.3-2$ \kms~ at 20-1000 K) at all radii. This may be the result from the fact that our model is 1-dimensional, or from the possible role of rotation in the line broadening. We are not able to test the latter effect without mapping W43-MM1 at higher spatial resolution. However large rotational velocities ($>$ 10 \kms) are likely to be found at radii closer than 600 AU \citep[][]{keto2010}, and moreover its main impact would be a small velocity shift of the molecular line profiles \citep[][]{belloche2002}. This high level of turbulence, derived from so many indicators, agrees with the conclusion of  \citet{plume1997} that regimes of massive-star formation are highly turbulent. Figure \ref{fig:turb} illustrates how the turbulent velocity varies with distance to the center of the object as derived from our model. We can then state that the turbulent motions increase with radius, in agreement with the turbulent core model of \citet{mckee2003} and Larson's studies in clouds \citep[][]{larson2003}. The lower degree of turbulence in the inner envelope can be due to the rotation and the density structure suppressing turbulence close to the protostar.

Additionally, Fig. \ref{fig:turb} shows the radii where the lines are mostly excited in our model (lines are dominated by one main turbulent velocity): lines with low $E_{up}$ are more sensitive to outer regions (i.e., higher energy level transitions come from the inner part). 

Moreover, the free-fall accretion rate would be around $6.4 \times 10^{-6}$ \msun yr$^{-1}$ (for $a_S\simeq0.3 $\kms) while here we detect ultrasonic gas motions and mass accretion rate between $10^{-4}$ and $10^{-2}$ \msun yr$^{-1}$ depending on the radius. Nevertheless, we stress that this very high derived accretion rate assumes spherical accretion, which may not be true. Also, if there is more than one central object in W43-MM1, the accretion rate per star could be considerably lower.
We can estimate the corresponding accretion luminosity, for the largest $\dot{M}_{acc}$ we derived, applying the following formula:
\begin{equation}
\label{Laccretion }
L_{acc}= \frac{G M_{\star} \dot{M}_{acc}}{R_{\star}}
\end{equation}
Considering a mass of 20 \msun within a radius of 100 R$_{\odot}$ for the protostar \citep[][]{hosokawa2010}, we get an accretion luminosity of the order of $3 \times 10^{4}$ \lsun. Comparing this value with the stellar luminosity of $10^5$ \Lsun~ derived by \citet{hosokawa2010} for a protostar with an accretion rate of only $10^{-3}$ \msun yr$^{-1}$ and 20 \Msun, we conclude that the derived accretion luminosity is unrealistic, compared to the observed total luminosity. As a consequence, the assumption of a single object might not be correct. However, \citet{hosokawa2010} also show that realistic accretion rates are much lower (by about one order of magnitude) than the values obtained from the simple formula we use, and as a consequence $\dot{M}$ might be more likely of the order of $10^{-3}$ \msun yr$^{-1}$ and the accretion luminosity of the order of a few $10^{3}$ \lsun, hence consistent with the observed total (stellar +accretion) luminosity. 

The derived accretion rate, although uncertain, is high enough (larger than $10^{-4}$ \msun yr$^{-1}$) to overcome the radiation pressure due to the star luminosity \citep[][]{mckee2003,yorke2008,hosokawa2009}. However, the protostar cannot reach the zero age main sequence by steady mass accretion if the accretion rate is so high \citep[][]{hosokawa2010}. Of course, one does not know if W43-MM1 will maintain such a high accretion rate over a long period of time. 

As recalled in the introduction, W43-MM1, though a very young source from indicators such as the SED, has already developed a hot core which is very likely at the origin of the high abundances of S-bearing species like H$_2$S and OCS \citep[][]{herpin2009} whose evaporation from grain surfaces could be important at such temperatures. Also, \citet{herpin2009} underlined the possibility of strong shocks occurring in the inner part (where $T>$100 K). Moreover, our study shows that the water abundance in W43-MM1 (in the outer envelope) is larger than in more evolved massive objects such as W3IRS5 and G31.41, which suggests fast desorption from grains. Hence, W43-MM1 seems not to fit in the assumed evolutionary diagram. Actually, there is some indication that high water abundances are not clearly linked to luminosity, mass, temperature or assumed evolutionary stage of the sources \citep[][]{marseille2010}. On the contrary, the high water abundance might be related to the presence of a HMC in the object. But, if so, one has to explain how high temperatures in the inner region can boost water production in the outer region. Another option for high water production in the inner regions is by shocks. As revealed by the violent gas motions detected in the envelope, W43-MM1 exhibits a huge infall and large turbulent velocities. Infall is generally associated with strong and fast outflows \citep[e.g.][]{klaassen2011,liu2011}. As a consequence, rather than being due to the presence of a HMC, the higher water abundance might be due to the presence of fast gas motions (i.e., shocks) as seen in low-mass objects \citep[][]{codella2010,nisini2010, kristensen2010} where \water~ abundances up to $10^{-5}$ can be found along outflow walls.

\section{Conclusions}
\label{sec:Conclusions}

We have presented here Herschel-HIFI observations of W43-MM1, for the first time, of as many as fourteen far-IR water lines (\water, \watersept, \waterhuit), CS(11--10), and C$^{18}$O(9--8) lines. We first analyzed the line profiles, then we modeled the observations using a radiative transfer code. We estimated outflow and infall velocities, turbulent velocity, and molecular abundances. 

Compared to what is observed towards low- and intermediate-mass objects, the water lines observed towards W43-MM1 are obviously brighter, but reveal the same broad and medium velocity components. The narrow component, from the hot core seen in W43-MM1, is not so commonly detected, likely because of opacity reasons. Hence observations of higher-energy water transitions are needed to probe this hot core.

From the modeling we estimate an outer water abundance of 8 $10^{-8}$. This value is higher than observed in other sources and could be due to the fast outflow and infall observed in W43-MM1. Moreover, the high level of turbulence derived in this source seems to be the usual regime in massive protostars, while the huge infall, i.e., mass accretion rate, very likely makes W43-MM1 a special case. Both turbulence and accretion rate vary with radius. If confirmed, the measured high level of turbulence combined with a high accretion rate agrees with the turbulent core model, although we cannot yet rule out without high spatial resolution data the competitive accretion model. Finally, the estimated accretion luminosity is high enough to overcome the expected radiation pressure.

HIFI mapping of the outflow, to be presented in a next paper, will help to determine the detailed kinematics of this component. In addition, the gas cooling budget will be studied in detail through a complete census of all water, plus [OI] and OH lines with PACS.

\begin{acknowledgements}
This program was made possible thanks to the HIFI guaranteed time. HIFI has been designed and built by a consortium of
institutes and university departments from across Europe, Canada and the
United States under the leadership of SRON Netherlands Institute for Space
Research, Groningen, The Netherlands and with major contributions from
Germany, France and the US. Consortium members are: Canada: CSA,
U.Waterloo; France: CESR, LAB, LERMA, IRAM; Germany: KOSMA,
MPIfR, MPS; Ireland, NUI Maynooth; Italy: ASI, IFSI-INAF, Osservatorio
Astrofisico di Arcetri- INAF; Netherlands: SRON, TUD; Poland: CAMK, CBK;
Spain: Observatorio Astron«omico Nacional (IGN), Centro de Astrobiolog«\u0131a
(CSIC-INTA). Sweden: Chalmers University of Technology - MC2, RSS \&
GARD; Onsala Space Observatory; Swedish National Space Board, Stockholm
University - Stockholm Observatory; Switzerland: ETH Zurich, FHNW; USA:
Caltech, JPL, NHSC.
HIPE is a joint development by the Herschel Science Ground
Segment Consortium, consisting of ESA, the NASA Herschel Science Center, and the HIFI, PACS and
SPIRE consortia.
Astrochemistry in Leiden is supported by the Netherlands Research School for Astronomy (NOVA), by a Spinoza grant and grant 614.001.008 from the Netherlands Organisation for Scientific Research (NWO), and by the European CommunityÕs Seventh Framework Program FP7/2007Ð2013 under grant agreement 238258 (LASSIE). We also thank the French Space Agency CNES for financial support. This research used the facilities of the Canadian Astronomy Data Centre (program m03bu45) operated by the the National Research Council of Canada with the support of the Canadian Space Agency. The James Clerk Maxwell Telescope is operated by the Joint Astronomy Centre on behalf of the Science and Technology Facilities Council of the United Kingdom, the Netherlands Organisation for Scientific Research, and the National Research Council of Canada. We thank C. Vastel, D. Johnstone and J. Mottram for their useful comments and PRISMAS team for useful discussions on absorption features.
 \end{acknowledgements}

\bibliography{biblio}
\bibliographystyle{aa}

\begin{appendix}

\section{Absorption lines on the line-of-sight: cold foreground gas ?}
\label{sec:absorption}

\begin{table*}
  \caption{Opacity, total column density and o$/$p ratio derived for each line-of-sight absorption component detected.}\label{los-W43MM1}
\begin{center}
\begin{tabular}{lccccc} \hline \hline
  & \multicolumn{5}{c}{l.o.s. component} \\ 
 Velocity (km$/$s)  & 67.2 & 71.0 & 79.5 & 82.5 & 87.8   \\ 
\hline                        
   $\tau_{557}$ & 1.00 $\pm$ 0.07 & 0.58 $\pm$ 0.06 & 1.71 $\pm$ 0.06 & 0.54 $\pm$ 0.06 & 0.7 $\pm$ 0.2 \\
   $\tau_{1113}$ & 0.74 $\pm$ 0.07 & 0.46 $\pm$ 0.09 & 1.5 $\pm$ 0.1 & 0.7 $\pm$ 0.1 & 0.09 $\pm$ 0.02 \\
   $\tau_{1669}$ & 1.2 $\pm$ 0.1 & 0.9 $\pm$ 0.1 & 3.10 $\pm$ 0.04 & 1.2 $\pm$ 0.1 & 0.28 $\pm$ 0.06  \\
\hline                        
   N ($10^{11}$ cm$^{-2}$) & 8.3 $\pm$ 0.9 & 4.9 $\pm$ 0.9 & 18.6 $\pm$ 0.6 & 7.2 $\pm$ 0.8 & 1.5 $\pm$ 0.3 \\
\hline                        
   o$/$p & 2.8 $\pm$ 0.6 & 2.5 $\pm$ 0.9  & 3.0 $\pm$ 0.2  & 2.6 $\pm$ 0.6  & 4.4 $\pm$ 1.9  \\
  \hline 
\end{tabular}
\end{center}
\end{table*}

\begin{figure}
\centering
\includegraphics[width=9cm, angle=0]{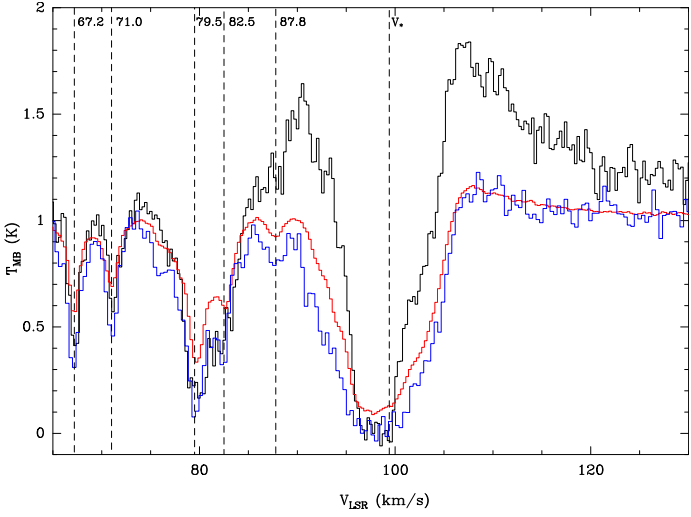}
\caption{Overplotted spectra (divided by the continuum) of the ground-state ortho- (black) and para-\water~ (red) and the \hoN (blue) lines. The different absorption components, apart from the main \water~ emission from the object, are shown by a dashed line and the corresponding velocity is given. The spectra have been smoothed to 0.3 \kms.}
\label{fig:absorption}
\end{figure}

The ground-state ortho- and para-\water~ and the \hoN~ spectra (see Fig.\ref{fig:absorption}) reveal the presence of deep absorptions at several velocities (67.2, 71.0, 79.5, 82.5, and 87.8 \kms; derived from Gaussian fits). The parameters of the absorption features are listed in Table \ref{los-W43MM1}. For each velocity component we derive the optical depth at the maximum absorption dip, based on the assumption that the excitation temperature is negligible with respect to the continuum brightness temperature. The optically thickest absorption is at 79.5 \kms~ ($\tau_{1669} \geq 3.1$), followed by components at 67.2 and 82.5 \kms~ while the two other absorptions are optically thinner. Because the absorption of the continuum signal at 79.5 \kms~ is almost complete, the absorbing material must be cold. The Gaussian fits reveal that the 67.2, 71.0 and 87.8 \kms~ absorptions are 1.4-1.9 \kms~ wide while the absorption at 79.5 \kms~ is definitely broader (5.3-5.9 \kms).

Assuming that all water molecules are in the ortho- and para ground states and that the excitation temperature is negligible with respect to the continuum brightness temperature, we calculate the column densities $N$ from the opacities applying the following formula:
\begin{equation}
\label{ }
N = \frac{g_l}{g_u} \frac{8\pi \tau \delta \varv \nu^3}{c^3 A}
\end{equation}
  
where $\tau$ is the optical depth, $\nu$ the line frequency, $\delta \varv$ the width, $c$ the speed of light, $A$ the Einstein coefficient, and $g_l$ and $g_u$ the degeneracy of the lower and upper levels of the transition, respectively. 

Low column densities are derived, of the order of a few $10^{11}$ cm$^{-2}$, except for the 79.5 \kms component where $N$ is several times larger (these values of the column density explain why the absorption are not seen for other  transitions). From these calculations we derive the o$/$p ratio for each component. Surprisingly, if the absorptions are due to cold foreground clouds, the o$/$p ratio does not strongly diverge from 3 if we take into account the uncertainties. This means that, in these clouds, water is produced in what is called the {\em high temperature regime} (T$_{spin} >$ 50K). Of course, we stress that opacities might be underestimated because of the assumptions we made.

\end{appendix}



%
%

\end{document}